\documentclass[final,3p,times,twocolumn]{elsarticle}

\usepackage{amsmath,amssymb,graphicx,color,psfrag,textcomp}

\usepackage[normalem]{ulem}

\newcommand\diff{\mathrm{d}}
\renewcommand\vec[1]{\boldsymbol{\mathrm{#1}}}
\newcommand\dotprod{\boldsymbol{\cdot}}
\newcommand\expect[1]{\left\langle\vphantom{\big(}#1\right\rangle}

\journal{Non-crystalline Solids}

\newcommand\e{\text{e}}
\renewcommand\i{\text{i}}
\DeclareMathOperator\Imag{Im}
\DeclareMathOperator\Real{Re}

\begin{document}

\begin{frontmatter}



\title{Space-resolved dynamics of a tracer in a disordered solid}


 \author[erlangen,lmu]{Thomas Franosch}
\author[erlangen]{Markus Spanner}
\author[lmu]{Teresa Bauer}
\author[erlangen]{Gerd E. Schr\"oder-Turk}
\author[oxford,mpi_stuttgart]{Felix H\"of{}ling}
\address[erlangen]{Institut f\"ur Theoretische Physik, Friedrich-Alexander-Universit\"at
Erlangen-N\"urnberg, Staudtstra{\ss}e 7, 91058 Erlangen, Germany}
\address[lmu]{Arnold Sommerfeld Center for Theoretical Physics and Center for
Nano Science (CeNS), Fakult{\"a}t f{\"u}r Physik,
Ludwig-Maximilians-Universit{\"a}t M{\"u}nchen, Theresienstra{\ss}e 37, 80333
M{\"u}nchen, Germany}
\address[oxford]{Rudolf Peierls Centre for Theoretical Physics, 1~Keble Road,
Oxford OX1 3NP, England, United Kingdom}
\address[mpi_stuttgart]{Max-Planck-Institut f\"ur Metallforschung,
Heisenbergstra{\ss}e 3, 70569
Stuttgart and Institut f\"ur Theore\-tische und Angewandte Physik, Universit\"at
Stuttgart, Pfaffenwaldring 57, 70569 Stuttgart, Germany
}
\begin{abstract}
\noindent
The dynamics of a tracer particle in a glassy matrix of obstacles displays slow complex transport as the free volume approaches a
critical value and the void space falls apart.
We investigate the emerging subdiffusive motion of the test particle by extensive molecular dynamics simulations
and characterize the spatio-temporal transport in terms of two-time correlation functions, including the time-dependent diffusion coefficient as well as the wavenumber-dependent
intermediate scattering function. We rationalize our findings within the framework of  critical phenomena and compare our data to a dynamic scaling
theory.
\end{abstract}


\end{frontmatter}



\section{Introduction}

\noindent
The structural dynamics in glass-forming liquids slows down by orders of magnitude  upon cooling or compression eventually leading
to a quasi-arrested state or a disordered solid.
Conventionally the dynamics close to the glass transition slows down uniformly as manifested in the divergence of a single time scale
characterizing the slowest process in the
system, referred to as the $\alpha$-process.
A coherent theoretical picture for a variety of phenomena has emerged by the mode-coupling theory (MCT) of the glass
transition~\cite{Goetze:Complex_Dynamics},  developed by Wolfgang G\"otze and collaborators in the last 25 years. In particular,
this  approach has explained how a two-step relaxation process with a nontrivial two-time fractal results from the
equations of motion close to a bifurcation point. The mechanism encoded in these equations to yield power laws with exponents that are not simple fractions
has no precedence in other areas of physics and yet has been identified as generic in nonlinear integro-differential equations.

The mathematical properties of the MCT solutions are well understood for single
component liquids~\cite{Goetze:1995}
and mixtures~\cite{Franosch:2002}; in particular, it has been shown that the solutions allow a representation in terms of a
continuous distribution of relaxation rates~\cite{Franosch:1999}. The success of the mode-coupling theory is based
on the fact that it not only  explains how the slow dynamics emerges in principle, but that it also has provided a series of
testable predictions which challenge the physical intuition obtained so far. To mention just
a few, MCT coherently explains the physics of  suspensions of hard-sphere colloids~\cite{Megen:1993,Megen:1998,Sperl:2005},
the reentrant phenomenon in 'attractive colloidal glasses'~\cite{Pham:2002,Dawson:2000}, a pronounced minimum in the light-scattering spectra
of supercooled liquids~\cite{Li:1992,Franosch:1997b,Singh:1998}, or the composition dependence of the structural relaxation in
mixtures~\cite{Foffi:2003,Goetze:2003}.
For many years only the standard mode coupling equations have been discussed and attempts to go beyond
have been focused on extension for non-spherical molecules~\cite{Schilling:1997,Franosch:1997,Schilling:2002,Chong:2002} and more
recently to include shear in colloidal suspensions~\cite{Fuchs:2002,Brader:2008}.

The overall success of MCT is encouraging to investigate more complex glass-transition scenarios to obtain a deeper insight
 into the nature and quality of the approximations involved.
A particularly interesting candidate are strongly size-disparate mixtures where structural arrest may occur in several steps.
There, the majority component of large particles can undergo a glass transition characterized by a frozen disordered structure,
whereas the minority species of smaller size meanders through the emerging network of channels. Neutron scattering experiments
and molecular dynamics simulations indicate a
significant separation of time scales for sodium silicate melts  also at finite concentration~\cite{Meyer:2004}, which could
be rationalized within standard MCT calculations~\cite{Voigtmann:2006} later. A similar splitting of relaxation times
has been found also for size-disparate soft spheres~\cite{Moreno:2006b} and Yukawa mixtures~\cite{Kikuchi:2007}. The dynamics within
a frozen matrix or nanoporous medium has been studied only recently~\cite{Kurzidim:2009,Kim:2009} corroborating an intriguing interplay of
several dynamic transitions predicted within an extension  of the mode-coupling theory for the dynamics within a
disordered matrix~\cite{Krakoviack:2005,Krakoviack:2007}.
The standard MCT of mixtures appears to qualitatively describe many aspects of the splitting of the dynamics~\cite{Voigtmann:2009},
yet ultimately predicts that the structural arrest
of the majority and minority particles occurs at the same critical point. The observed peculiarities  are rationalized as a precursor
phenomenon for the two-step freezing not contained in the standard MCT. Krakoviack's extensions to disordered matrices treats
the frozen obstacles and the fluid on unequal footing~\cite{Krakoviack:2005,Krakoviack:2009} thereby allowing for multiple transitions.
Although both approaches appear to give a satisfactory qualitative picture, theoretical issues remain  that are poorly understood. First,
the freezing in the disordered solid is accompanied by a divergent length scale suggesting that the transition is driven by
an entire hierarchy of length scales rather than the Lindemann length for caging. It appears that MCT in its current form is
ill-suited to deal with this phenomenon~\cite{Goetze:Complex_Dynamics}. This manifests itself in 'spurious long-time anomalies'~\cite{Krakoviack:2009} that,
if treated correctly, lead to a change in the exponent of the anomalous transport observed in the mean-square displacement~\cite{Papenkort:2010}.

An alternative approach dealing with the dynamics of a single tracer in a
disordered matrix is within the framework
of critical phenomena. The localization transition of the tracer is due to an underlying percolation transition of the
void space as the density of the matrix is altered. As is well known, this geometric problem leads to a self-similar
distribution of clusters~\cite{Stauffer:Percolation} entailing a series of scaling predictions. It has been suggested
that the dynamics of the Lorentz model shares the same universality class with
a random resistor network with power-law distributed conductances~\cite{Straley:1982,Machta:1985,Halperin:1985}.
The critical  dynamics of the Lorentz model in the vicinity of the percolation threshold can be described within a
scaling ansatz for the van Hove correlation function~\cite{Kertesz:1983}. Furthermore, corrections to scaling due to irrelevant scaling variables are not negligible and have been included recently~\cite{Lorentz_PRL:2006,Percolation_EPL:2008}.

In this paper we investigate the Lorentz model by extensive computer simulations and show
that the mean-square displacement  becomes anomalous with an exponent differing from a simple fraction.
The vanishing of the diffusion coefficient upon approaching the localization transition is connected to
 the subdiffusive motion and the divergence of the correlation length
or mean cluster size. We discuss, in particular the motion of tracers that are confined to the percolating cluster in terms of  the mean-square displacement,
as well as the time-dependent diffusion coefficient averaged over all initial conditions of the tracers. We then elucidate the transport as response to an
alternating external field in terms of the frequency-dependent conductivity and susceptibility.
Furthermore, we characterize the space-resolved dynamics in terms of the
intermediate scattering function, which is accessible to scattering experiments in principle. Its long-time limit, known
as the Lamb-M\"o{\ss}bauer factor or nonergodicity factor, reveals the trapping of particles inside the finite clusters.

\section{The Lorentz model}

\begin{figure}[t]
\includegraphics[width=\linewidth]{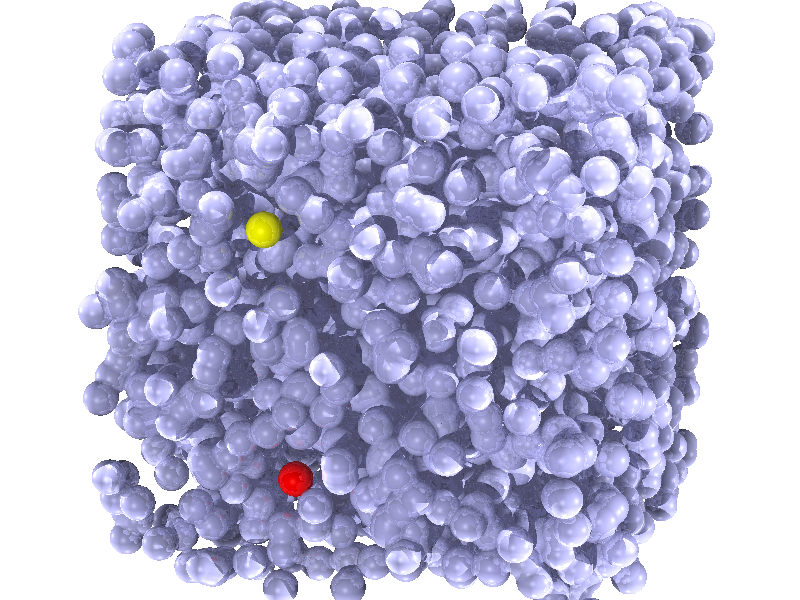}
\caption{Typical configuration for a random array of obstacles where the void space for a tracer is close to the percolation threshold. The tracer particles
(highlighted) displayed
have equal radius as the obstacles and are confined to the void space.
}
\label{fig:porous_blob}
\end{figure}

\noindent
A simple model for transport in disordered solids capturing all the relevant ingredients for complex transport was
introduced by Lorentz already in 1905~\cite{Lorentz:1905}. There a single
tracer or equivalently a system of particles which do not interact among themselves traverses a course of
immobilized obstacles. The position  of the tracer is excluded from the space occupied by the obstacles thus confined to the void space.
In the simplest case  hard spherical obstacles  are distributed independently and randomly in space and
consequently the number of obstacles $N$  per volume $V$ characterizes the emerging structure completely.
Then a dimensionless control parameter $n^* =  N\sigma^d/V$ can be introduced,
where $\sigma$ is the distance of the hard-core exclusion between the tracer and a single obstacle.
Note that for independent obstacles, only $\sigma$ determines the void space allowing to
consider the tracer as pointlike and the obstacles as having a radius $\sigma$. Then the obstacles naturally overlap
resulting in clusters of excluded space. Alternatively, one may view the
obstacles as pointlike and the tracer of radius $\sigma$, reminiscent of a random Galton board.
Figure~\ref{fig:porous_blob} displays the case where the obstacles and the tracer are of the same size, with the highlighted particles moving through the frozen maze.

A self-consistent mode-coupling kinetic theory has been introduced by G\"otze, Leutheusser, and Yip~\cite{Goetze:1981a,Goetze:1981b,Goetze:1982}
which provides a theory for various aspects of the dynamics of the Lorentz gas. In particular, their approximation
gives an extremely accurate value for the first order density correction~\cite{Weijland:1968} to the Lorentz-Boltzmann
value for the diffusion coefficient and correctly reflects the long-time tails in the velocity autocorrelation function
(VACF)~\cite{Ernst:1971a}. Furthermore, they correctly predict a localization transition at a critical obstacle
density $n_c^*$ at which anomalous transport occurs. The competition between the
critical law of the localization and the long-time tail then leads to density-dependent apparent exponents, a prediction
that has been verified only recently by computer simulations~\cite{Lorentz_LTT:2007}. Although the kinetic theory of
the Lorentz model gives a remarkable description for low and moderate densities, the regime very close  to the localization is more involved.

The immediate vicinity of the arrest of transport is characterized by a
diverging length scale describing
the distribution of clusters of the void space. Already at moderate densities the void space falls apart
into disconnected finite components, hence confining the motion of the tracer particles.
Thus even below the critical
density, the system becomes non-ergodic in the sense that tracers initially placed in such a pocket remain there
 forever. Long-range transport occurs only on the infinite void space cluster percolating through the array of obstacles.
At the transition point this infinite cluster ceases to exist and all tracer particles are trapped. Thus the origin
 of the localization is the underlying geometric percolation transition of
the void space. Percolation itself has been the investigated for many years and a coherent picture has been
 established~\cite{Stauffer:Percolation}, summarized below.

Similar to a continuous phase transition, percolation exhibits a diverging length
scale $\xi \sim |\epsilon|^{-\nu}$
which corresponds here to the linear dimension of the largest finite pocket.
Here the reduced obstacle density $\epsilon = (n^*- n_c^*)/n_c^*$
quantifies the distance to the critical density and plays the role of a
separation parameter.
The
 weight
of the infinite void space
cluster also vanishes as a power law,  $P_\infty \sim  (-\epsilon)^{\beta}$ for
$\epsilon < 0$, as
the obstacle density is increased.
 Directly at $n_c^*$, the infinite void space becomes fractal with a fractal
dimension $d_\text{f} = d-\beta/\nu$, and in coexistence with the infinite cluster, there is a self-similar hierarchy of finite clusters.
In contrast to thermal phase transitions a second length scale, viz. the
mean-square cluster size~$\ell$,
plays an important role for the percolation transition. Using scaling arguments, one can show that its divergent behavior is determined
 by the same two exponents as
 $\ell \sim |\epsilon|^{-\nu + \beta/2}$.
The critical exponents for percolation are known from
computer simulations~\cite{Stauffer:Percolation,Percolation_EPL:2008} and attain the approximate
 values $\nu \approx 0.88$ and $\beta \approx 0.41$ in three-dimensional space, $d=3$.

\section{Motion on the infinite cluster}

\noindent
Let us discuss first the dynamics of tracer particles that move on the
percolating cluster only.
Transport on the percolating cluster is expected to become fractal
precisely at the critical obstacle density. After an elapsed time~$t$ a particle has moved
typically a distance $t^{1/d_\text{w}}$, where $d_\text{w}$ is referred to as
the walk dimension and attains the value of $d_\text{w} \approx 4.81 $~\cite{Percolation_EPL:2008} in
the three-dimensional Lorentz model. For the mean-square displacement $\delta
r^2_\infty(t) := \expect{[\vec{R}(t)- \vec{R}(0)]^2}_\infty$
 this  implies $\delta r_\infty^2(t) \sim
t^{2/d_\text{w}}$, where the index $\infty$ indicates that averaging is only for
particles initially located on the infinite cluster. Below the critical obstacle
density, transport should follow this subdiffusive behavior for intermediate times and then cross over
to
a diffusive motion $\delta r^2_\infty = 6 D_\infty t$ for length
scales larger than $\xi \sim (-\epsilon)^{-\nu}$, where the cluster appears
homogeneous.
 Both aspects can be combined in the scaling law
\begin{equation}
 \delta r^2_\infty(t;\epsilon) = t^{2/d_\text{w}}
\delta \tilde{r}_\infty^2(\tilde{t}) \, , \qquad \tilde{t} = t/t_x
\end{equation}
i.e., all mean-square displacements have the same shape in a double-logarithmic
representation. The dependence on the reduced obstacle density $\epsilon$ is
solely by a change of scale $t_x = t_x(\epsilon)$.
 The scaling
function $\delta \tilde{r}^2_\infty(\cdot)$ attains a constant for short
rescaled times $\tilde{t} \ll 1$, reflecting the anomalous transport. The
length scale where the crossover to diffusion occurs is given by the
correlation length, which then in turn determines the crossover time to
$t_\infty \sim \xi^{d_\text{w}} \sim |\epsilon|^{-\nu d_\text{w}}$.
The crossover to diffusion is recovered by
imposing $\delta\tilde{r}^2_\infty(\tilde{t}) \sim \tilde{t}^{1-2/d_\text{w}}$
 at long rescaled times $\tilde{t}\gg 1$. As a consequence  the corresponding
diffusion
coefficient vanishes as a power law, $D_\infty \sim
t_x^{2/d_\text{w}-1}\sim (-\epsilon)^{\nu(d_\text{w}-2)}$. Since the
particles on the infinite cluster are the only ones that contribute to
long-range motion, one derives the scaling behavior of the diffusion coefficient
for an all-cluster-average by $D(\epsilon) = P_\infty(\epsilon)
D_\infty(\epsilon)\sim (-\epsilon)^{\mu}$ with
the conductivity exponent $\mu = \beta+ \nu(d_\text{w}-2)$.

We have performed extensive computer simulations for the Lorentz gas with
ballistic particles as tracers. In this case the trajectory in configuration space
consists of a series of
straight lines connecting the scattering events with the obstacles. We have
employed deterministic specular scattering which in particular conserves the
magnitude of
the velocity $v = |\vec{v}(t)|$.
The trajectories are generated using the standard event-driven algorithm already
used by Bruin~\cite{Bruin:1972} combined with the usual
method~\cite{Frenkel:MD,Glassy_GPU:2010} for calculating correlation
functions online, optimized for exponentially large time
scales. The  infinite cluster has been identified by a standard Voronoi
tessellation to generate trajectories for the infinite-cluster-only averages.

\begin{figure}
\includegraphics[width=\linewidth]{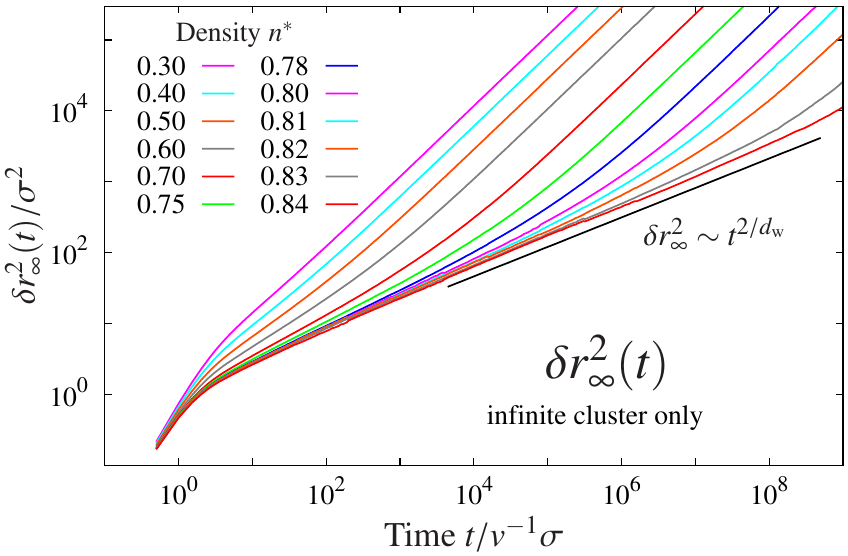}
\caption{Double-logarithmic representation of the mean-square displacements
$\delta r_\infty^2(t)$  for particles   on the infinite cluster.
Obstacle density increases from top to bottom. The straight line indicates
anomalous transport with a walk dimension of $d_\text{w} = 4.81$.
}
\label{fig:msd_infinite_cluster}
\end{figure}
The simulation results for the mean-square displacement $\delta r^2_\infty(t)$
are displayed in Fig.~\ref{fig:msd_infinite_cluster} for various obstacle
density below the localization transition. All curves show ballistic motion
$\delta r^2_\infty(t) = v^2 t^2$ for times smaller than the
inverse collision rate or mean collision time $\tau_c$.
In the low-density regime, $n\to 0$,
the collision time grows, $\tau_c \sim 1/n \sigma^{d-1} v$, yet close to the
percolation transition, the density
dependence can be ignored. For low-obstacle densities the motion crosses over
directly to a linear increase of $\delta r_\infty^2(t)$ in time, as is expected
for normal transport. As the critical density $n_c^*$ is approached a window of
subdiffusive transport emerges and the mean-square displacements increase as a
power law in accordance with  theoretical prediction. Directly at the
critical point $n_c^* \approx 0.839$ the power law $t^{2/d_\text{w}}$ is
observed over more than 6 six decades in time. The long-time behavior for all
densities below $n_c^*$ is again diffusive but sets in at later and later times
as $\epsilon \uparrow 0$. A short inspection of the curves reveals that the
shapes of the crossover functions are similar and scaling behavior is anticipated
to hold on a semi-quantitative level. Yet including the corrections to
scaling~\cite{Percolation_EPL:2008}
should corroborate the validity of the scaling hypothesis more generally.
A more detailed analysis of the scaling
behavior of $\delta r^2_\infty(t)$ can be found in~\cite{Spanner:thesis}.

\section{All-cluster-averaged motion}

\noindent
Conventionally the Lorentz model considers averages over all allowed starting
positions of the tracer, i.e., both on any of the finite clusters as well as on
the infinite cluster. Long-range transport occurs only on the percolating
cluster, but close to the threshold the finite clusters become arbitrarily
large and contribute to transport at all scales.
The motion of an ensemble of tracers in such a fractal landscape is also
self-similar in the time-domain introducing a new exponent $z$ characteristic
of the subdiffusive
motion at the critical point. The unconstrained mean-square
displacement is defined as $\delta r^2(t) := \expect{[\vec{R}(t)- \vec{R}(0)]^2}$, where the
average includes initial positions in the infinite and in finite clusters according
to their weight as well as over the disorder.
It increases again as a  power law,
$\delta r^2(t) \sim t^{2/z}$, where the dynamic exponent $z$ is larger than the
walk dimension $d_\text{w}$. Since the all-cluster averaged mean-square
displacement has already been discussed  in Refs.\ \citealp{Lorentz_PRL:2006} and~\citealp{Lorentz_JCP:2008},
 here we focus on the time-dependent diffusion coefficient
\begin{equation}\label{eq:def_D_t}
 D(t) := \frac{1}{d}\expect{[\vec{R}(t)- \vec{R}(0)] \dotprod \vec{v}(0)}\,,
\end{equation}
i.e., the correlation of the time-dependent displacement  $\vec{R}(t)- \vec{R}(0)$ with the initial velocity $\vec{v}(0) =\dot{\vec{R}}(0)$.
The definition of $D(t)$ holds
in any dimension $d$, in particular  for  $d=3$. Because of  time-translational invariance
the time-dependent diffusion coefficient is related to the mean-square displacement via
\begin{equation}
D(t) = \frac{1}{2d} \frac{\diff }{\diff t} \delta r^2(t) \, .
\end{equation}
Similarly, the velocity autocorrelation function $Z(t) := (1/d)\expect{\vec{v}(t) \dotprod \vec{v}(0)}$  encodes the time-dependent motion via
$ D(t) = \int_0^t Z(t')\, \diff t'$ consistent with the Green-Kubo relation for the diffusion constant $D := D(t\to \infty)$ as the long-time limit of the
time-dependent diffusion coefficient.

\begin{figure}
\includegraphics[width=\linewidth]{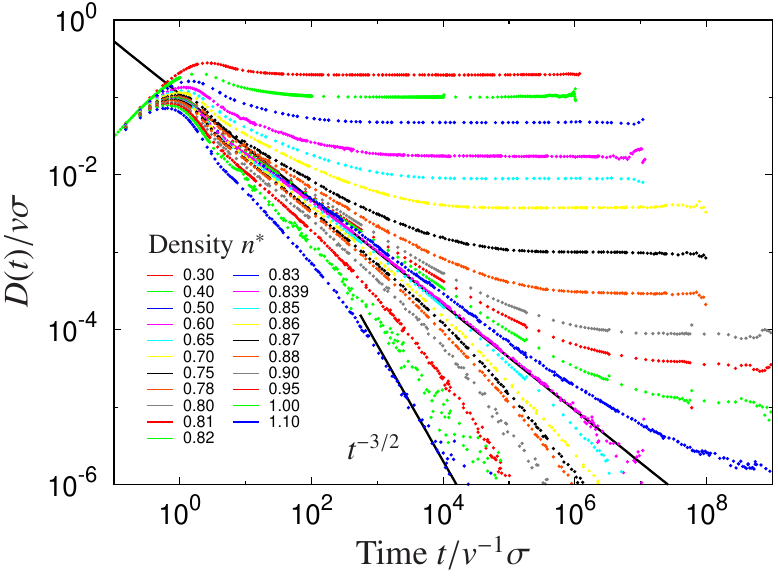}
\caption{Double-logarithmic representation of the time-dependent diffusion coefficient $D(t) = (1/6)\diff \delta r^2(t) /\diff t$. Density increases from top to
bottom. The  full line indicates subdiffusive behavior $D(t) \sim t^{2/z-1}$ at the critical obstacle density $n_c \approx 0.839$. The thick line indicates
the power law $t^{-3/2}$ characterizing the long-time decay in the localized phase.
}
\label{fig:D_t}
\end{figure}

The simulation data for the time-dependent diffusion coefficient $D(t)$ are displayed in Fig.~\ref{fig:D_t}. At short times, $D(t) = v^2 t/d$
 increases linearly and the curves for all scatterer densities superimpose. This reflects the free flight of the tracer until the first scattering event
occurs at the collision time $\tau_c$.
At  time scales of the order of the
collision time, the time-dependent diffusion coefficient displays a maximum
due  to the cage
effect familiar from the zero crossing in the velocity autocorrelation function.
For longer times, $D(t)$ decreases monotonically, effects of inertia
become less and less relevant, and the structural relaxation dominates the transport.

For low and moderate obstacle densities, $D(t)$ rapidly saturates at its long-time limit $D(t\to \infty)$, which corresponds to the diffusion constant $D$ of the tracer.
Upon increasing the number of scatterers towards the percolation threshold, diffusion is rapidly suppressed. In this regime  $D(t)$ displays a crossover from
power-law decrease to saturation and one anticipates that scaling holds.
A detailed analysis of the  scaling properties of the three-dimensional Lorentz model in terms of the mean-square displacement can be found in Ref.~\citealp{Lorentz_PRL:2006};
the scaling of the time-dependent diffusion coefficient for a two-dimensional Lorentz model is discussed in Ref.~\citealp{Lorentz_2D:2010}.

Precisely at the percolation threshold, $D(t) \sim t^{2/z-1}$ is expected to
hold for $t\to \infty$ as is inferred from the corresponding subdiffusive behavior of the mean-square displacement. Our simulations display such anomalous transport
over more than five decades in time allowing the determination of the exponent $z=6.25$, consistent with the prediction that the universality class of the three-dimensional
Lorentz model coincides with the one of a random resistor network with power-law distributed
weak conductances~\cite{Lorentz_PRL:2006,Lorentz_JCP:2008}.

In the high-density regime, $n^*> n^*_c$, the time-dependent diffusion coefficient vanishes in the long-time limit reflecting the fact that all tracers are trapped in finite
pockets of the void space. Again, close to criticality the data follow the anomalous behavior until a crossover time and then decreases more rapidly. Yet
the long-time
behavior is not characterized by an exponential decay since power law distributed exit rates from cul-de-sacs lead to  long-time tails
in the localized phase~\cite{Machta:1985}. For the three-dimensional case, the velocity autocorrelation function (VACF) is predicted to decay as $t^{-5/2}$ which corresponds to $t^{-3/2}$ for the time-dependent diffusion coefficient
as indicated in Fig.~\ref{fig:D_t}. To the best of our knowledge, this algebraic decay has not been observed before.

\section{Frequency-dependent conductivity}\label{sec:conductivity}

\noindent
In the context of ion conductors one is interested in the frequency-dependent complex conductivity $\sigma(\omega)$ which quantifies the
alternating electric current density $\vec{j}(\omega)$ as  response to
an frequency-dependent  homogeneous electric field $\vec{E}(\omega)$ in the linear regime as $\vec{j}(\omega) = \sigma(\omega)\, \vec{E}(\omega)$.
For disordered materials that are statistically isotropic the current is parallel to the electric field and the conductivity $\sigma(\omega)$ transforms as a scalar.
A decomposition into real and imaginary parts, 
$\sigma(\omega) =\Real[\sigma(\omega)] + \i \Imag[\sigma(\omega)]$,
reveals that the
in-phase component $\Real[\sigma(\omega)]\geq 0$
describes the loss due to friction,
whereas  $\Imag[\sigma(\omega)]$ encodes the storage of energy in the system.
For the important case that the conducting ions can be viewed as independent,
the conductivity is obtained by $\sigma(\omega) = q^* n_{\text{ion}}
\mu(\omega)$, where $q^*$ is the effective charge of the ions, $n_\text{ion}$ their
number density, and  $\mu(\omega)$ the frequency-dependent mobility.

The linear response theorem extends the well-known Einstein relation
$D = \mu k_B T$ to finite frequencies $Z(\omega) = k_B T \mu(\omega)$, where
the frequency-dependent generalization $Z(\omega)$ of the diffusion coefficient
$D$ is given by
\begin{equation}
 Z(\omega) = \frac{1}{d} \int_0^\infty \expect{\vec{v}(t) \dotprod \vec{v}(0)}
\exp(\i \omega t) \diff t \, ,
\end{equation}
i.e., as the one-sided Fourier transform of the velocity autocorrelation
function.
In particular, the Green-Kubo relation for the diffusion coefficient is recovered
for the stationary case, $\omega=0$. By partial integration one obtains a representation of $Z(\omega)$ in terms of
the time-dependent diffusion coefficient
\begin{equation}\label{eq:Z_omega}
 Z(\omega) = D - \text{i} \omega \int_0^\infty [ D(t) - D]  \text{e}^{\text{i} \omega t} \diff t\, ,
\end{equation}
with the diffusion coefficient $D= D(t\to \infty)$.

\begin{figure}
\includegraphics[width=\linewidth]{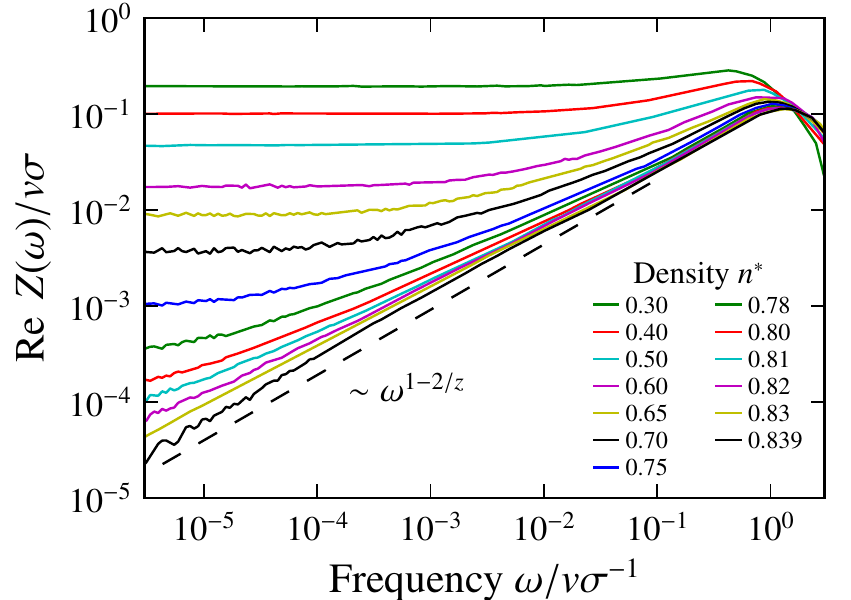}
\caption{Frequency-dependent diffusion coefficient
 $\Real[Z(\omega)]$ on  double-logarithmic scales for various obstacle
densities below the localization threshold. Densities increase from top to
bottom. The broken line indicates the power law $\omega^{1-2/z}$ expected for
 anomalous transport at the critical density.  }
\label{fig:Z_omega}
\end{figure}

Figure~\ref{fig:Z_omega} displays the real part of the frequency-dependent
diffusion coefficient, $\Real[Z(\omega)]$, obtained via a numerical one-side Fourier transform from our simulation data according to Eq.~\ref{eq:Z_omega}.
At the critical density $n_c^*$, the power-law dependence $\Real[Z(\omega)]
\sim \omega^{1-2/z}$ holds over almost five decades in frequency. The
fractal behavior of $Z(\omega)$ is inherited from an algebraic
long-time decay of of the VACF, $Z(t) \sim t^{2/z-2}$, or equivalently from the
subdiffusive increase of the  mean-square displacement, $\delta r^2(t) \sim
t^{2/z}$.

For obstacle densities below $n_c^*$, the curves attain a finite value
at low-frequencies which is identified with the diffusion constant $D=
Z(\omega=0)$. At high frequencies, the frequency-dependent diffusion coefficient
approaches the critical law. The two regimes of anomalous transport and
diffusion merge at a characteristic crossover frequency $\omega_x$ which is shifted
to lower values as the critical density is approached.

These  observations suggest the scaling behavior
\begin{equation}
 Z(\omega;\epsilon) = (-\i \omega)^{1-2/z} \mathcal{Z}_{\pm}(\hat{\omega}) \,
, \qquad \hat{\omega} = \omega/\omega_x\,,
\end{equation}
in the close vicinity of the localization transition.
Here the scaling functions $\mathcal{Z}_{+}(\cdot)$ and $\mathcal{Z}_-(\cdot)$ refer to
densities above ($+$) and below ($-$) the critical density,
respectively. For large rescaled frequencies $\hat{\omega} \gg 1$, the tracers
cannot explore the entire landscape and transport is fractal,
$\mathcal{Z}_\pm(\hat{\omega} \to \infty) \to const$, i.e., indistinguishable from
the critical point.  In particular, the constant is the same on both sides of the
transition.

On the conducting side, $\mathcal{Z}_-(\hat{\omega}) \sim (-\i
\hat{\omega})^{2/z-1}$ for $\hat{\omega} \to 0$ is enforced by the requirement
of a finite conductivity, which in turn implies the scaling law $D(\epsilon)
\sim \omega_x^{1-2/z}$. Defining the conductivity exponent $\mu$ via
$D(\epsilon) \sim (-\epsilon)^{\mu}$ yields $\omega_x \sim (-\epsilon)^{\mu
z/(z-2)}$.

On the insulating side, it is advantageous to discuss the complex polarizability
$\chi(\omega) = \sigma(\omega)/(-\i \omega)$ rather than the conductivity
itself. Let us define the susceptibility $\tilde{\chi}(\omega) = Z(\omega)/(-\i \omega)$
by dividing out factors that are irrelevant for the present discussion, $\chi(\omega) = (q^* n_\text{ion}/k_B T) \tilde{\chi}(\omega)$. Then
$\tilde{\chi}(\omega)$ has the dimension  of the square of a length.
A finite polarizability is obtained by imposing $\mathcal{Z}_+(\hat{\omega})
\sim (-\i \hat{\omega})^{2/z}$ for low frequencies $\hat{\omega} \ll 1$.
The static susceptibility diverges upon approaching the critical obstacle
density $\tilde{\chi}(\omega=0,\epsilon) \sim \omega_x^{-2/z}$. Because of  geometrical reasons
this susceptibility should  diverge in the same way as the mean-square cluster size
$\tilde{\chi}(\omega=0,\epsilon)\sim \ell^2 \sim \epsilon^{-(2\nu-\beta)}$.  Combining
the preceding arguments reveals the scaling relation between the conductivity
exponent and dynamic  exponent $z$ to $z = (2\nu-\beta+\mu)/(\nu-\beta/2)$. Then one also observes that the crossover frequency scales as $\omega_x \sim t_x^{-1}$, i.e
 the inverse of the
crossover time $t_x$ of the mean-square displacement for the infinite cluster only.

\begin{figure}
\includegraphics[width=\linewidth]{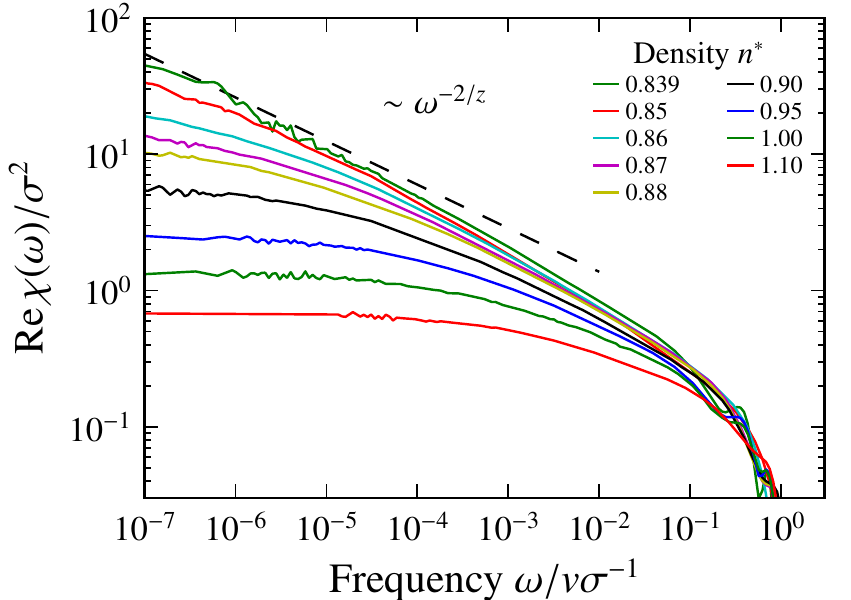}
\caption{Frequency-dependent susceptibility
 $\Real[\tilde{\chi}(\omega)]$ on  double-logarithmic scales for various obstacle
densities above the localization threshold. Densities increase from top to
bottom. The broken line indicates the power-law divergence $\omega^{-2/z}$ expected for
 anomalous transport at the critical density. }
\label{fig:susceptibility}
\end{figure}

The simulation results for the susceptibility on the insulating side are displayed in Fig.~\ref{fig:susceptibility}. As discussed above,
the real part becomes finite in the static case reflecting the finite polarizability of the system. Approaching the delocalization transition by diluting the obstacles
gives rise to a strong increase of the static polarizability. At the critical density, the susceptibility displays power-law behavior in the frequency with finite
response only for alternating external  electric fields.

\section{Self-intermediate scattering function}

\noindent
Spatio-temporal information on the dynamics of the tracer can be extracted by
considering the self-intermediate scattering function (ISF) defined by
\begin{equation}
 F_s(q,t) := \expect{\!\exp(\i\,\vec{q}\dotprod\Delta \vec{R}(t)} \, .
\end{equation}
Thus $F_s(q,t)$ is the characteristic function of the displacements
$\Delta \vec{R}(t) := \left[ \vec{R}(t)-\vec{R}(0)\right]$ considered as the random variable. We have
anticipated already that due to statistical isotropy $F_s(q,t)$ does not depend on
the direction of the wavevector $\vec{q}$, but only on its magnitude $q =
|\vec{q}|$. The intermediate scattering function encodes all moments of the
displacement via derivatives, for instance the mean-square displacement can be
obtained as $\delta r^2(t) = - d \lim_{q\to 0} \partial^2 F_s(q,t)/\partial
q^2$. Scattering techniques such as neutron-spin echo or photon correlation
spectroscopy have direct access to the ISF and there $\hbar \vec{q}$  plays the role
of the momentum transfer from the tracer to the neutrons or photons.
\begin{figure}
\includegraphics[width=\linewidth]{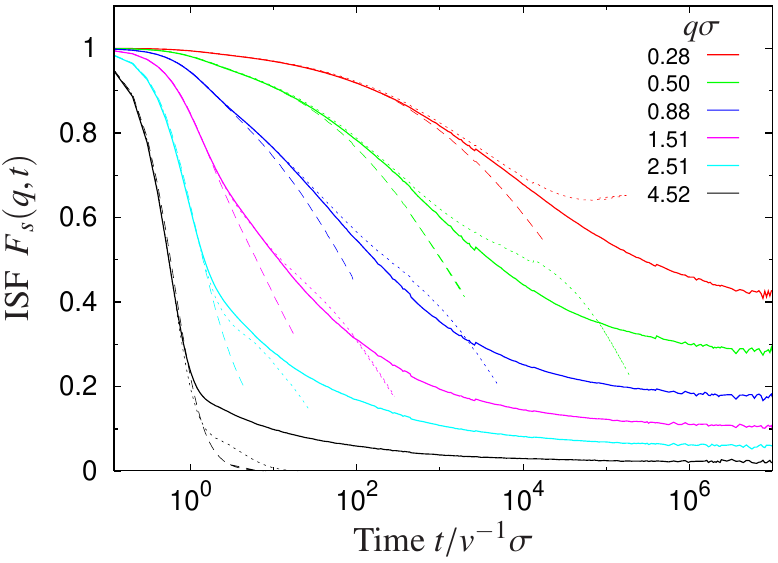}
\caption{Self-intermediate scattering function $F_s(q,t)$ on a logarithmic time
 scale at the critical density $n_c^* =0.839$. Wavenumbers increase from  top to
bottom. The full lines correspond to the simulation data, the broken ones to the Gaussian approximation $F_s^{\text{Gauss}}(q,t)$, and the dotted ones include the first non-Gaussian correction $F_s^\text{NGP}(q,t)$.}
\label{fig:Fq_t}
\end{figure}
The simulation results for the ISF for small to moderate wavenumbers $q$ for
obstacle density directly at $n_c^*$ are  displayed in Fig.~\ref{fig:Fq_t}. As
expected transport becomes slower on larger length scales. In contrast to glassy
dynamics characterized by a two-step relaxation process, here no plateau develops
at intermediate times. Rather all intermediate scattering function appear to
saturate at some finite level. Since the infinite cluster has non-extensive
weight at the critical point~\cite{Stauffer:Percolation} its dynamics does not contribute to the ISF shown.
Yet the infinite hierarchy of finite clusters coexisting with the percolation
cluster leads to non-ergodic behavior as manifested in the ISF, see below. Since the
finite clusters are present even below the critical density, one should in fact
expect non-ergodic behavior due to the falling apart of the configuration space
at all finite densities.

The spatio-temporal information encoded in the ISF is revealed by gradually changing
 the  wavenumber $q$. To highlight the non-trivial behavior we have
included the Gaussian approximation $(d=3)$
\begin{equation}
 F_s^{\text{Gauss}}(q,t) = \exp\left(- q^2 \delta r^2(t)/6 \right),
\end{equation}
in Fig.~\ref{fig:Fq_t}. This approach ignores all cumulants except for  the second, which
reduces to the mean-square displacement $\delta r^2(t)$. The
Gaussian model accounts approximately for the increase of time scale upon changing the
wavenumber and also provides a quantitative description of the initial decay of
the ISF. Interestingly the Gaussian model becomes worse on increasing the
length scale,  demonstrating the fact that transport at the critical density
never becomes Gaussian even at the largest scales. A similar observation  on the
planar Lorentz model has been made recently in the
context of fluorescence correlation spectroscopy~\cite{FCS_Scaling:2010}.
Including also the first non-Gaussian parameter (NGP),
\begin{equation}
\alpha_2(t) :=  \frac{3}{5}
\expect{\left[\Delta \vec{R}(t)^2\right]^2} \bigg/ \expect{\Delta \vec{R}(t)^2 }^2 -1,
\end{equation}
in the intermediate scattering function,
\begin{equation}
 F_s^{\text{NGP}}(q,t) = \e^{- q^2 \delta r^2(t)/6} \left\{ 1 + \frac{1}{2} \alpha_2(t)\left[ \frac{q^2 \delta r^2(t)}{6} \right]^2 \right\} \, ,
\end{equation}
describes the simulation data even to times larger by a factor of 10. Nevertheless, this approximation too fails on a qualitative level for long times since it predicts
a decay of all ISFs to zero due to the unbounded increase of the mean-square displacement.

\section{Lamb-M{\"o}ssbauer factor}

\noindent
The long-time limit of the intermediate scattering function
\begin{equation}
 f_s(q) := \lim_{t\to\infty} F_s(q,t)\, ,
\end{equation}
is referred to as Lamb-M{\"o}ssbauer factor or non-ergodicity parameter, also known as the Edwards-Anderson parameter in the spin glass community~\cite{Goetze:Complex_Dynamics}. A non-vanishing $f_s(q)$ indicates
that dynamic correlations are persistent forever implying that the dynamics is non-ergodic. Roughly speaking $f_s(q)$ measures the fraction of particles that
are trapped on length scale $2\pi/q$.
 For the Lorentz model the presence of finite clusters at all densities
suggests that the non-ergodicity transition has to be disentangled from the localization transition. A clear distinction can be made by considering the long-wavelength
limit $f_s(0) = \lim_{q\to0} f_s(q)$ of the Lamb-M\"ossbauer factor, since $1-f_s(0) = P(\epsilon)$ where $P(\epsilon)$ denotes the weight of the percolating cluster.
Since above  the localization threshold $n^* > n_c^*$ the infinite cluster ceases to exist, $f_s(0)$ attains the value unity, whereas increasing the obstacle density from
below leads to an expected critical behavior $1-f_s(0)  \sim (-\epsilon)^\beta$.

For finite wavenumbers this sharp transition is rounded off and crossover scaling should hold~\cite{Kertesz:1983}. A convenient form is obtained by postulating
$1-f_s(q;\epsilon) = |\epsilon|^{\beta} \hat{f}_\pm(q \xi)$, where the scaling functions $\hat{f}_\pm$ refer to densities above the threshold (+) and the below the
localization transition (-).
The critical behavior of the long-wavelength limit is recovered by requiring $\hat{f}_-(x\to 0) =const.$, and
$\hat{f}_+(x\to 0) = 0$, respectively. For large wavenumbers and densities close to the transition the dynamics cannot resolve the finiteness of the correlation
length and should become independent of the separation parameter, which is achieved by imposing $\hat{f}_\pm(x) \sim x^{\beta/nu}$. This in turn implies a
singular dependence of the Lamb-M\"ossbauer factor on the wavenumber $1-f_s(q;\epsilon=0) \sim q^{\beta/\nu}$ directly at the threshold.

\begin{figure}[t]
\includegraphics[width=\linewidth]{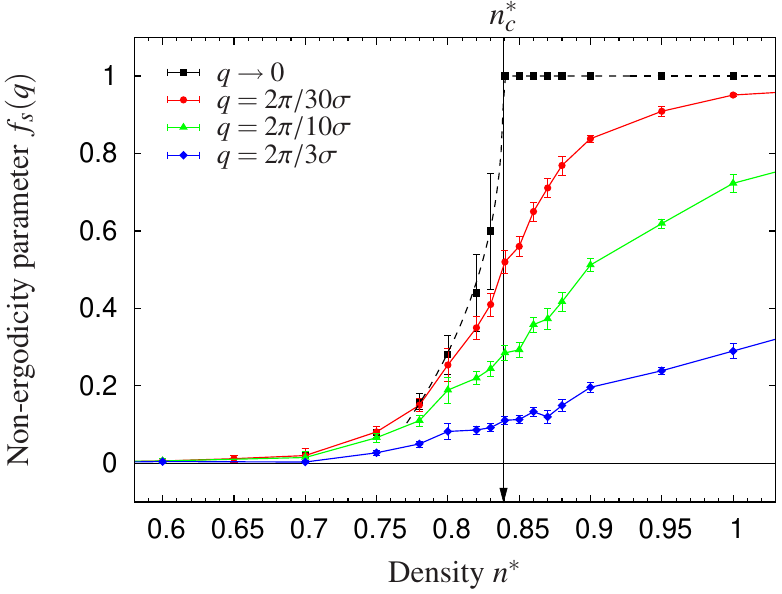}
\caption{Lamb-M\"o{\ss}bauer factor $f_s(q) = F_s(q,t\to \infty)$ as a function of
the obstacle density $n^*$. As the wavenumber decreases the curves steepen. The
extrapolation to zero wavenumber exhibits a sharp transition at the critical
obstacle density $n_c^*$. The broken line corresponds to a power law $(-\epsilon)^\beta$ and serves as a guide to the eye.}
\label{fig:Nonergodicity}
\end{figure}

The Lamb-M\"ossbauer factors obtained from our simulation results for the ISF as extrapolation to infinite
times are displayed in Fig.~\ref{fig:Nonergodicity}. Upon increasing the density $f_s(q)$ increases monotonically
 and the shape develops into a rapid crossover for small wavenumbers.
Yet for finite wavenumber nothing peculiar happens at the percolation threshold. Only the extrapolation to the
 long wavelength limit indicates the localization transition
in the sense that it converges to  unity rapidly as $n_c^*$ is approached from below. Due to the large statistical
uncertainties we have not attempted to validate the
scaling law. Nevertheless the approach of the long-wavelength limit towards unity is compatible with the power
 law expected for the weight of the infinite cluster.



\section{Summary and Conclusion}

Transport in disordered material as studied in terms of the Lorentz model displays many facets that go beyond the
simple subdiffusive increase of the mean-square displacement. We have elucidated some of these aspects as they are manifest in the mean-square displacement for
tracers that are confined to the percolating cluster only,
 the time-dependent diffusion coefficient for an all-cluster average, the frequency-dependent conductivity and susceptibility in response to an \emph{AC} electric field, as
well as the intermediate scattering function and its long-time limit, the Lamb-M\"ossbauer factor.

The percolation transition of the underlying
void space induces a self-similar geometric fractal with anomalous transport at the localization threshold. The interplay of time and length scales
leads to a striking behavior of the complex transport properties many aspects of which can be rationalized in terms of scaling laws. The scaling predictions
require three independent exponents as input and we have demonstrated that all three values can be revealed by considering different limits of certain two-time correlation
functions. In particular, the detailed study shows that simplified concepts such as fractional Brownian motion or continuous time random walks (CTRW) cannot explain
the dynamics as soon as   spatio-temporal information beyond the mean-square displacement at the threshold is studied.

The localization transition in the Lorentz model in its simplest variant has now been studied from various aspects and nice agreement of extensive computer simulations  with theoretical notions
has been achieved. An open question is how
anomalous transport occurs in a class of glass-forming liquids that displays a strong separation of time scales, and a possible route to an answer is
to include more microscopic details into
the Lorentz model, thereby gradually adding more complexity.
These changes could include studying the motion of a tracer in a frozen  matrix of \emph{correlated} obstacles generated from a
snapshot of a real liquid. In particular, one has to clarify the role of the narrow gaps that emerge in the frozen labyrinth and their influence on the value of the
dynamic or conductivity exponent.
Furthermore, it is interesting
 to investigate the effect of interacting tracers inside the quasi-arrested host matrix since for long times the tracers will strongly interact
in the ramified network of channels. A further interesting aspect is   the interaction of a dynamic maze on the particle, i.e., the effect of
vibrations of the slow component on the dynamics of the tracer.

\section{Acknowledgments}

\noindent
Thomas Franosch is indebted to  Wolfgang G\"otze for many years of  discussions
and his priceless insight into complex transport connected to the glass and
localization transition.

Financial support  from the Deutsche Forschungsgemeinschaft via contract No. FR
850/6-1 is gratefully acknowledged.
This project is supported by the  German Excellence Initiative via the program ``Nanosystems Initiative Munich (NIM).''


\begin{thebibliography}{10}

\bibitem{Goetze:Complex_Dynamics}
W.~G\"otze, Complex Dynamics of Glass-Forming Liquids -- A Mode-Coupling
  Theory, Oxford, 2009.

\bibitem{Goetze:1995}
W.~G{\"o}tze, L.~Sj\"ogren, General properties of certain nonlinear
  integrodifferential equations, J. Math. Anal. Appl. 195 (1995) 230--250.

\bibitem{Franosch:2002}
T.~Franosch, T.~Voigtmann, {Completely monotone solutions of the mode-coupling
  theory for mixtures}, {J. Stat. Phys.} {109} (2002) {237}.

\bibitem{Franosch:1999}
T.~Franosch, W.~G\"otze, {Relaxation rate distributions for supercooled
  liquids}, {J. Phys. Chem. B} {103}~({20}) (1999) {4011--4017}.

\bibitem{Megen:1993}
W.~van Megen, S.~M. Underwood, Dynamic-light-scattering study of glasses of
  hard colloidal spheres, Phys. Rev. E 47~(1) (1993) 248--261.

\bibitem{Megen:1998}
W.~van Megen, T.~C. Mortensen, S.~R. Williams, J.~M\"uller, Measurement of the
  self-intermediate scattering function of suspensions of hard spherical
  particles near the glass transition, Phys. Rev. E 58~(5) (1998) 6073--6085.

\bibitem{Sperl:2005}
M.~Sperl, Nearly logarithmic decay in the colloidal hard-sphere system, Phys.
  Rev. E 71~(6) (2005) 060401.

\bibitem{Pham:2002}
K.~N. Pham, A.~M. Puertas, J.~Bergenholtz, S.~U. Egelhaaf, A.~Moussaid, P.~N.
  Pusey, A.~B. Schofield, M.~E. Cates, M.~Fuchs, W.~C.~K. Poon, Multiple glassy
  states in a simple model system, Science 296~(5565) (2002) 104--106.

\bibitem{Dawson:2000}
K.~Dawson, G.~Foffi, M.~Fuchs, W.~G\"otze, F.~Sciortino, M.~Sperl,
  P.~Tartaglia, T.~Voigtmann, E.~Zaccarelli, Higher-order glass-transition
  singularities in colloidal systems with attractive interactions, Phys. Rev. E
  63~(1) (2000) 011401.

\bibitem{Li:1992}
G.~Li, W.~Du, X.~Chen, H.~Cummins, N.~Tao, {Testing Mode-coupling Predictions
  for alpha-Realaxation and beta-Relaxation in Ca0.4K0.6(NO3)1.4 near the
  Liquid-Glass Transition by Light-Scattering}, {PHYSICAL REVIEW A} {45}~({6})
  (1992) {3867--3879}.

\bibitem{Franosch:1997b}
T.~Franosch, W.~G\"otze, M.~R. Mayr, A.~P. Singh, Evolution of structural
  relaxation spectra of glycerol within the gigahertz band, Phys. Rev. E 55~(3)
  (1997) 3183--3190.

\bibitem{Singh:1998}
A.~Singh, G.~Li, W.~G{\"o}tze, M.~Fuchs, T.~Franosch, H.~Cummins, Structural
  relaxation in orthoterphenyl: a schematic-mode-coupling-theory-model
  analysis, J. Non-Cryst. Solids 235-237 (1998) 77.

\bibitem{Foffi:2003}
G.~Foffi, W.~G\"otze, F.~Sciortino, P.~Tartaglia, T.~Voigtmann, Mixing effects
  for the structural relaxation in binary hard-sphere liquids, Phys. Rev. Lett.
  91~(8) (2003) 085701.

\bibitem{Goetze:2003}
W.~G\"otze, T.~Voigtmann, Effect of composition changes on the structural
  relaxation of a binary mixture, Phys. Rev. E 67~(2) (2003) 021502.

\bibitem{Schilling:1997}
R.~Schilling, T.~Scheidsteger, Mode coupling approach to the ideal glass
  transition of molecular liquids: Linear molecules, Phys. Rev. E 56~(3) (1997)
  2932--2949.

\bibitem{Franosch:1997}
T.~Franosch, M.~Fuchs, W.~G\"otze, M.~Mayr, A.~Singh, {Theory for the
  reorientational dynamics in glass-forming liquids}, {Phys. Rev. E} {56}~({5})
  (1997) {5659--5674}.

\bibitem{Schilling:2002}
R.~Schilling, Reference-point-independent dynamics of molecular liquids and
  glasses in the tensorial formalism, Phys. Rev. E 65~(5) (2002) 051206.

\bibitem{Chong:2002}
S.-H. Chong, W.~G\"otze, Idealized glass transitions for a system of dumbbell
  molecules, Phys. Rev. E 65~(4) (2002) 041503.

\bibitem{Fuchs:2002}
M.~Fuchs, M.~E. Cates, Theory of nonlinear rheology and yielding of dense
  colloidal suspensions, Phys. Rev. Lett. 89~(24) (2002) 248304.

\bibitem{Brader:2008}
J.~M. Brader, M.~E. Cates, M.~Fuchs, First-principles constitutive equation for
  suspension rheology, Physical Review Letters 101~(13) (2008) 138301.

\bibitem{Meyer:2004}
A.~Meyer, J.~Horbach, W.~Kob, F.~Kargl, H.~Schober, Channel formation and
  intermediate range order in sodium silicate melts and glasses, Phys. Rev.
  Lett. 93~(2) (2004) 027801.

\bibitem{Voigtmann:2006}
T.~Voigtmann, J.~Horbach, Slow dynamics in ion-conducting sodium silicate
  melts: {S}imulation and mode-coupling theory, Europhys. Lett. 74~(3) (2006)
  459--465.

\bibitem{Moreno:2006b}
A.~J. Moreno, J.~Colmenero, Anomalous dynamic arrest in a mixture of large and
  small particles, Physical Review E (Statistical, Nonlinear, and Soft Matter
  Physics) 74~(2) (2006) 021409.

\bibitem{Kikuchi:2007}
N.~Kikuchi, J.~Horbach, Mobile particles in an immobile environment: Molecular
  dynamics simulation of a binary yukawa mixture, EPL 77 (2007) 2600.

\bibitem{Kurzidim:2009}
J.~Kurzidim, D.~Coslovich, G.~Kahl, Single-particle and collective slow
  dynamics of colloids in porous confinement, Phys. Rev. Lett. 103~(13) (2009)
  138303.

\bibitem{Kim:2009}
K.~Kim, K.~Miyazaki, S.~Saito, Slow dynamics in random media: Crossover from
  glass to localization transition, EPL 88~(3) (2009) 36002.

\bibitem{Krakoviack:2005}
V.~Krakoviack, Liquid-glass transition of a fluid confined in a disordered
  porous matrix: A mode-coupling theory, Phys. Rev. Lett. 94~(6) (2005) 065703.

\bibitem{Krakoviack:2007}
V.~Krakoviack, Mode-coupling theory for the slow collective dynamics of fluids
  adsorbed in disordered porous media, Phys. Rev. E 75~(3) (2007) 031503.

\bibitem{Voigtmann:2009}
T.~Voigtmann, J.~Horbach, Double transition scenario for anomalous diffusion in
  glass-forming mixtures, Phys. Rev. Lett. 103~(20) (2009) 205901.

\bibitem{Krakoviack:2009}
V.~Krakoviack, Tagged-particle dynamics in a fluid adsorbed in a disordered
  porous solid: Interplay between the diffusion-localization and liquid-glass
  transitions, Phys. Rev. E 79~(6) (2009) 061501.

\bibitem{Papenkort:2010}
S.~Papenkort, F.~H\"ofling, T.~Franosch, T.~VoigtmannIn preparation.

\bibitem{Stauffer:Percolation}
D.~Stauffer, A.~Aharony, Introduction to Percolation Theory, 2nd Edition,
  Taylor \& Francis, London, 1994.

\bibitem{Straley:1982}
J.~P. Straley, Non-universal threshold behaviour of random resistor networks
  with anomalous distributions of conductances, J. Phys. C 15 (1982)
  2343--2346.

\bibitem{Machta:1985}
J.~Machta, S.~M. Moore, Diffusion and long-time tails in the overlapping
  {L}orentz gas, Phys. Rev. A 32 (1985) 3164.

\bibitem{Halperin:1985}
B.~I. Halperin, S.~Feng, P.~N. Sen, Differences between lattice and continuum
  percolation transport exponents, Phys. Rev. Lett. 54 (1985) 2391--2394.

\bibitem{Kertesz:1983}
J.~Kert{\'e}sz, J.~Metzger, Properties of the density relaxation function in
  classical diffusion models with percolation transition, J.~Phys.~A 16 (1983)
  L735--L739.

\bibitem{Lorentz_PRL:2006}
F.~H{\"o}f\/ling, T.~Franosch, E.~Frey, Localization transition of the
  three-dimensional {L}orentz model and continuum percolation, Phys. Rev. Lett.
  96 (2006) 165901.

\bibitem{Percolation_EPL:2008}
A.~Kammerer, F.~H{\"o}f\/ling, T.~Franosch, Cluster-resolved dynamic scaling
  theory and universal corrections for transport on percolating systems, EPL 84
  (2008) 66002.

\bibitem{Lorentz:1905}
H.~A. Lorentz, Le mouvement des electrons dans les metaux, Arch. N{\'e}erl.
  Sci. Exact Natur. 10 (1905) 336--370.

\bibitem{Goetze:1981a}
W.~G{\"o}tze, E.~Leutheusser, S.~Yip, Dynamical theory of diffusion and
  localization in a random, static field, Phys. Rev. A 23 (1981) 2634.

\bibitem{Goetze:1981b}
W.~G{\"o}tze, E.~Leutheusser, S.~Yip, Correlation functions of the hard-sphere
  {L}orentz model, Phys. Rev. A 24 (1981) 1008.

\bibitem{Goetze:1982}
W.~G{\"o}tze, E.~Leutheusser, S.~Yip, Diffusion and localization in the
  two-dimensional {L}orentz model, Phys. Rev. A 25 (1982) 533.

\bibitem{Weijland:1968}
A.~Weijland, J.~M.~J. van Leeuwen, Non-analytic density behaviour of the
  diffusion coefficient of a {L}orentz gas {II}, Physica (Amsterdam) 38 (1968)
  35.

\bibitem{Ernst:1971a}
M.~H. Ernst, A.~Weijland, Long time behaviour of the velocity auto-correlation
  function in a {L}orentz gas, Phys. Lett. A 34 (1971) 39.

\bibitem{Lorentz_LTT:2007}
F.~H{\"o}f\/ling, T.~Franosch, Crossover in the slow decay of dynamic
  correlations in the {L}orentz model, Phys. Rev. Lett. 98~(14) (2007) 140601.

\bibitem{Bruin:1972}
C.~Bruin, Logarithmic terms in the diffusion coefficient for the {L}orentz gas,
  Phys. Rev. Lett. 29~(25) (1972) 1670--1674.

\bibitem{Frenkel:MD}
D.~Frenkel, B.~J. Smit, Understanding Molecular Simulation, 2nd Edition,
  Academic Press, London, 2001.

\bibitem{Glassy_GPU:2010}
P.~H. Colberg, F.~H{\"o}f{}ling, Accelerating glassy dynamics on graphics
  processing units, arXiv:0912.3824 [physics.comp-ph] (2009).

\bibitem{Spanner:thesis}
M.~Spanner, Transport in the correlated lorentz model, Master's thesis,
  Friedrich-Alexander-Universit\"at Erlangen N\"urnberg (2010).

\bibitem{Lorentz_JCP:2008}
F.~H{\"o}f\/ling, T.~Munk, E.~Frey, T.~Franosch, Critical dynamics of ballistic
  and {B}rownian particles in a heterogeneous environment, J. Chem. Phys.
  128~(16) (2008) 164517.

\bibitem{Lorentz_2D:2010}
T.~Bauer, F.~H{\"o}f{}ling, T.~Munk, E.~Frey, T.~Franosch, Localization
  transition in the two-dimensional {L}orentz model, arXiv:1003.2918
  [cond-mat.soft].

\bibitem{FCS_Scaling:2010}
F.~H{\"o}f{}ling, K.-U. Bamberg, T.~Franosch, Anomalous transport resolved in
  space and time by fluorescence correlation spectroscopy, arXiv:1003.3762
  [cond-mat.soft].

\end{thebibliography}







\end{document}